# High-pressure x-ray diffraction study of SrMoO$_4$ and pressure-induced structural changes

Daniel Errandonea[1,*], Ravhi S. Kumar[2], Xinghua Ma[3], and Chaoyang Tu[3]

[1] Departamento de Física Aplicada-ICMUV, Universitat de València, Edificio de Investigación, c/Dr. Moliner 50, 46100 Burjassot, Valencia, Spain

[2] High Pressure Science and Engineering Center, Department of Physics and Astronomy, University of Nevada, 4505 Maryland Parkway, Las Vegas, Nevada 89154-4002, USA

[3] Fujian Institute of Research on the Structure of Matter, Chinese Academy of Sciences, Fuzhou, Fujian 350002, and Graduated School of Chinese Academy of Science, 100039 Beijing, China

**Abstract:** SrMoO$_4$ was studied under compression up to 25 GPa by angle-dispersive x-ray diffraction. A phase transition was observed from the scheelite-structured ambient phase to a monoclinic fergusonite phase at 12.2(9) GPa with cell parameters $a$ = 5.265(9) Å, $b$ = 11.191(9) Å, $c$ = 5.195 (5) Å, and $\beta$ = 90.9°, Z = 4 at 13.1 GPa. There is no significant volume collapse at the phase transition. No additional phase transitions were observed and on release of pressure the initial phase is recovered, implying that the observed structural modifications are reversible. The reported transition appeared to be a ferroelastic second-order transformation producing a structure that is a monoclinic distortion of the low-pressure phase and was previously observed in compounds isostructural to SrMoO$_4$. A possible mechanism for the transition is proposed and its character is discussed in terms of the present data and the Landau theory. Finally, the EOS is reported and the anisotropic compressibility of the studied crystal is discussed in terms of the compression of the Sr-O and Mo-O bonds.

PACS numbers: 62.50.+p, 61.50.Ks, 64.70.Kb

[*] Corresponding author, Email: daniel.errandonea@uv.es, Fax: (34) 96 3543146, Tel.: (34) 96 354 4475

## I. Introduction

Metal molybdates (AMoO$_4$) and tungstates (AWO$_4$) of relatively large bivalent cations (ionic radius > 0.99 Å; A = Ca, Ba, Sr, Pb or Eu, and Cd only for the molybdates) crystallize in the so-called scheelite structure (scheelite = CaWO$_4$) [1], which belongs to the tetragonal space group (SG) $I4_1/a$. The scheelite structure of strontium molybdate (SrMoO$_4$) is shown in Fig. 1. There, it can be seen that in this structure each Mo site is surrounded by four equivalent O sites in approximately tetrahedral symmetry about that site. On the other hand, each Sr site is surrounded by eight O sites in approximately octahedral symmetry.

Scheelite-structured compounds belonging to the molybdate and tungstate families possess attractive luminescence and interesting structural properties. Because of their physical and chemical properties they are used in several technological applications. In particular, scheelite-type crystals are used as scintillators [2], laser-host materials [3], cryogenic detectors for dark matter [4], or heterogeneous catalysts [5]. Alkaline-earth molybdates form part of these compounds. They produce green luminescence required for the uses of electro-optical devices [6] and recent studies have shown that these molybdate crystals have good prospects as possible negative electrode materials to replace the graphite presently being used in the Li-ion batteries [7]. On top of that, scheelite molybdates and tungstates are also being considered for the development of eye-safe Raman lasers [8]. Understanding the electro-optical and structural properties of these compounds is very important for the above mentioned applications.

High-pressure research has probed to be an efficient tool to improve the understanding of the main physical properties of AWO$_4$ and AMoO$_4$ compounds.

However, most of the previous studies have been performed on scheelite-type tungstates [9 – 20], having been established that they undergo a sequence of high-pressure phase transformations with space group changes $I4_1/a \rightarrow I2/a \rightarrow P2_1/n$. On top of that, optical absorption measurements [21] and luminescence studies [22] have shown that the electronic structure of scheelite tungstates is also strongly affected by pressure. In contrast to the tungstates, there are many open questions regarding the high-pressure structural behaviour of the molybdates. They have been studied under compression using Raman spectroscopy [23 – 27], and these studies reported evidence of structural changes at pressures ranging from 5 GPa to 12 GPa. Further, in $BaMoO_4$ and $CdMoO_4$ evidence of a second phase transition was found at 9 GPa and 25 GPa, respectively. Regarding high-pressure crystallographic studies, they have been performed in $CaMoO_4$ and $BaMoO_4$ up to about 15 GPa [28, 29]. These studies found that the x-ray diffraction patterns of the high-pressure phases of $CaMoO_4$ and $BaMoO_4$ can be fitted to a monoclinic fergusonite-type structure (SG: $I2/a$), resembling their high-pressure behaviour that of the isomorphic tungstates [11, 12, 15]. On the contrary, in the case of $CdMoO_4$ previous x-ray diffraction studies [30] suggested that the first pressure-driven phase transition is from the tetragonal $I4_1/a$ space group to the monoclinic $P2/c$ space group. On the other hand, contrasting with $BaMoO_4$ and $CdMoO_4$, in $SrMoO_4$ Raman measurements performed up to 37 GPa [25] only found a phase transition around 12 – 15 GPa. However, *ab initio* calculations predicted that more than one phase transition should take place in this pressure range [31]. In addition, they also suggested that the post-scheelite phase of $SrMoO_4$ is monoclinic but belongs to the space group $P2_1/n$. All these facts show that despite the experimental and theoretical efforts made in the past, we have not yet achieved a full understanding of the high-pressure structural behavior of scheelite-type molybdates.

The aim of the present study is to examine comprehensively the crystal stability of $SrMoO_4$ up to 25 GPa. In order to improve the current understanding of the structural behavior of scheelite-type $AMoO_4$ compounds we have performed angle dispersive x-ray powder diffraction (ADXRD) measurements in a diamond-anvil cell (DAC) at room temperature (RT). From our experiments we found that strontium molybdate undergoes a scheelite-to-fergusonite phase transition at 12.2 GPa, being this transition reversible. No other structural change was detected up to 25 GPa. In addition to that, based upon our experimental results and the Landau theory, we concluded that the reported phase transition is a second-order ferroelastic transformation. The reported results may be relevant for a better understanding of the structural behavior not only of scheelite-structured molybdates, but also of vanadates, germanates, and silicates.

**II. Experimental details**

The samples used in our experiments consisted in pre-pressed pellets prepared using a finely ground powder obtained from the single crystal of $SrMoO_4$. This crystal was grown at the Fujian Institute of Research by the Czochralski method [32]. The obtained crystal was optically transparent and color free. Before loading the samples into the DAC, their phase purity was tested with a laboratory x-ray diffractometer operated with Cu $K_\alpha$ radiation. The diffraction lines for $SrMoO_4$ collected at ambient pressure (0.0001 GPa) showed only a single scheelite phase with cell parameters $a$ = 5.394(5) Å and $c$ = 12.019(9) Å, in agreement with reported literature values [33].

$SrMoO_4$ powder samples were loaded in a 130 μm hole of rhenium gasket in a

Mao-Bell-type DAC with diamond culet sizes of 350 μm. A few ruby grains were also loaded with the sample for pressure determination [34] and silicone oil was used as pressure-transmitting medium [35, 36]. ADXRD experiments were carried out at Sector 16-IDB of the HPCAT, at the Advanced Photon Source (APS), with an incident wavelength of 0.41514 Å. The monochromatic x-ray beam was focused down to 10 × 10 μm$^2$ using Kickpatrick-Baez mirrors. The images were collected using a MAR345 image plate located 350 mm away from the sample and then integrated and corrected for distortions using FIT2D [37]. The structure solution, and refinements were performed using the POWDERCELL [38] program package.

### III. Results and Discussion

#### A. Low-pressure phase

The *in situ* ADXRD data measured at different pressures are shown in Fig. 2. The x-ray patterns could be indexed with the scheelite structure (stable at normal conditions) to 11.3 GPa. In Fig. 3, we show an x-ray diffraction pattern of SrMoO$_4$ measured at 0.5 GPa. The spectrum is plotted together with the refined structure model and the residuals of the refinement with the aim of illustrating the quality of the structural refinements used to extract the lattice parameters and bond lengths presented in this work. The full Rietveld refinement of the profile measured at 0.5 GPa with the scheelite model converged to small R-factors: $R_{WP}$ = 1.65 %, $R_P$ = 1.2 %, and $R(F^2)$ = 1.45 % (116 reflections). The starting model for the refinement was taken from the previously reported crystal structure of SrMoO$_4$ [33]. Table I summarizes the lattice parameters and atomic positions obtained for SrMoO$_4$ at 0.5 GPa. For every analyzed pressure up to 11.3 GPa, we found a good agreement between the refined profiles and the experimental diffraction patterns. We also observed that within the accuracy of the refinements the atomic positions remain nearly unchanged from ambient pressure to 11.3 GPa.

A splitting and broadening of the diffraction peaks is observed at 13.1 GPa together with the appearance of new reflections. In particular, the strongest intense peak of scheelite, the (112) reflection, observed around $2\theta = 7.7°$ at 11.3 GPa splits into two peaks. The same fact occurs with the (200) and (204) reflections located at 11.3 GPa around $2\theta = 9.1°$ and $2\theta = 12.4°$, respectively. In addition to that, at 13.1 GPa new weak peaks can be clearly observed in Fig. 2 at $2\theta = 4.2°$ and at $2\theta = 12.8°$. These facts are indicative of a structural phase transition around 12.2(9) GPa, which is in agreement with previous Raman observations [25] and with the transition pressure estimated from the $MoO_4$/Sr radii ratio [39].

From the refinement of the x-ray diffraction patterns measured up to 11.3 GPa, we extracted the pressure dependence of the lattice parameters, cell volume, and axial ratios for scheelite $SrMoO_4$. These results are summarized in Figs. 4 and 5. There it can be seen that as in other scheelite-structured molybdates [29] and tungstates [11, 12, 15] the compression of $SrMoO_4$ is highly anisotropic, being the *c*-axis more compressible than the *a*-axis. The *c/a* axial ratio decreases from 2.228 at ambient pressure to 2.152 at 11.3 GPa (see Fig. 4). This anisotropy in the axial compressibility of $SrMoO_4$ is comparable with that of $SrWO_4$ [11]. It should be mentioned that the *c/a* ratio for $SrMoO_4$ and $SrWO_4$ at ambient pressure is 2.228 and 2.206 [11], respectively. On top of that, the unit-cell volumes of the molybdates are generally smaller than those of the tungstate analogues [1], as can be seen by comparing the data shown in Fig. 5 with those published in Ref. [11]. The higher *c/a* ratio in molybdates can be explained by the higher

cation–cation electrostatic repulsion in SrMoO$_4$ than that in SrWO$_4$ [12]. The fact that in both compounds $c/a$ evolves in a similar way under compression is related to the fact that the compressibility of both SrMoO$_4$ and SrWO$_4$ is mainly due to the compression of the SrO$_8$ dodecahedra as we will show later in this section of the paper.

The pressure-volume curves shown in Fig. 5 were analyzed in the standard way using a third-order Birch-Murnaghan equation of state (EOS) [40]. The bulk modulus ($B_0$), its pressure derivative ($B_0$'), and the atomic volume ($V_0$) at zero pressure obtained for the scheelite phase of SrMoO$_4$ are $B_0$ = 71(3) GPa, $B_0$' = 4.2(4), and $V_0$ = 349.7(4) Å$^3$. The fitted EOS, shown as a solid line in Fig. 5, matches very well the experimental data. On top of that, the obtained bulk modulus agrees very well with the value calculated from elastic constant data ($B_0$ = 74 GPa) [25]. Previously it has been empirically determined that the bulk modulus in scheelite-structured ABX$_4$ compounds is related to the cation charge density of the AX$_8$ polyhedra [11, 41]. In particular, it has been established that $B_0$ (GPa) = *610 Z/d$^3$*, where Z is the cationic formal charge of A and $d$ is the mean A-X distance at ambient pressure in Å [11]. From our data we obtained an Sr-O distance of 2.578 Å, which agrees within 1% with the value reported by Gurmen [33]. Applying the formula given above to this value a bulk modulus of 71 GPa is obtained, which gives additional support to the fitted EOS.

From our experimental data, we have also investigated the evolution of cation-anion distances in SrMoO$_4$. According to our structural refinements, within the pressure stability range of the scheelite structure, the change of the oxygen position coordinates for SrMoO$_4$ under compression is insignificant, being comparable to the experimental errors. This observation is good agreement with the conclusions obtained from a single-crystal study carried out up to 4.1 GPa by Hazen *et al.* in other molybdates [41]. Such behavior can be expected for the scheelite type lattice, as the structure is made up of the hard $(MoO_4)^{2-}$ anions which are surrounded by the $Sr^{2+}$ ions to balance the charge. As with pressure the $(MoO_4)^{2-}$ ions do not show any noticeable changes, the oxygen positions do not change significantly with pressure. Based upon this observation, to calculate the pressure evolution of the cation-anion distances, we assumed at all pressures the atomic positions given in Table I. Fig. 6 shows the evolution of the atomic distances between nearest neighbors with increasing pressure. The decrease of Sr-O distances can be compared with the rigidity of the Mo-O bond. In Fig. 6 it can be seen that there are two Sr-O distances, the largest distance being more compressible than the shorter one. We also found that the second neighbors Mo-O distances also are more compressible than the first neighbors Mo-O distances. These conclusions provide support to the description of $AMoO_4$ scheelites in terms of nearly uncompressible anion-like $MoO_4$ tetrahedra surrounded by charge compensating cations. Upon compression the $MoO_4$ units remain essentially undistorted and the reduction of the unit-cell size is mainly accounted by the compression of the Sr-O dodecahedral environment. It is important to note that along the *a*-axis the $MoO_4$ units are directly aligned, whereas along the *c*-axis there is a Sr cation between each $MoO_4$ tetrahedra (see Fig. 1). Therefore, the different arrangement of hard $MoO_4$ tetrahedra along the *c*- and *a*-axes explains the anisotropic axial compressibility of $SrMoO_4$. On the other hand, the fact that the $MoO_4$ tetrahedra behave basically as uncompressible rigid units in comparison with the $SrO_8$ polyhedra is what allows the bulk modulus formula proposed in Ref. [11] to make accurate estimations of $B_0$ as shown above.

A comparison of the present results obtained for SrMoO$_4$ under high pressure with the reported structural variation at high temperature [42] of this compound shows a fairly close but inverse relation with each other. This fact is illustrated in Fig. 7, where we have plotted the volume dependence of the axial ratio of both compression and thermal expansion. A similar phenomenon was previously documented in CaMoO$_4$ [43]. The high-temperature studies on SrMoO$_4$ and CaMoO$_4$ [42, 43] shows a systematic increase in *c/a* with pressure as against the systematic decrease with pressure of the present investigation. In the particular case of SrMoO$_4$, the change of anisotropy with volume shows the same behavior under compression or expansion. This is also a consequence of the fact that the less rigid (higher ionic) Sr–O bonds have significant expansion or compression compared to the Mo–O. Therefore, the structural evolution under pressure or temperature is basically governed by the SrO$_8$ polyhedron. The crystal chemistry of SrMoO$_4$ at high temperature shows a similar linear dependence of the Sr–O bond with pressure or temperature, whereas the Mo–O bonds remain almost invariant in both situations. Thus, we conclude that in scheelite-type molybdates there is an inverse relationship between pressure and temperature, as previously documented for other ABO$_4$ compounds, like LaNbO$_4$ [44].

**B. High-pressure phase**

As we mentioned above, the ADXRD spectra of SrMoO$_4$ exhibit notorious changes between 11.3 GPa and 13.1 GPa (see Fig. 2). These changes are completely reversible upon pressure release. Beyond 11.3 GPa some of the diffraction peaks split and additional diffraction peaks emerge. In particular, the appearance of new peaks around $2\theta = 4.2°$ and $2\theta = 12.8°$ are clearly distinguishable. The observed splitting of peaks and the appearance of new reflections suggest the occurrence of a pressure-induced phase transition. At pressures higher than 13.1 GPa, some of the broadened diffraction peaks develop into two clearly separated diffraction peaks. In particular the splitting of the (112) and (200) is also easily identifiable in Fig. 2. The measured ADXRD patterns of the high-pressure phase can be indexed on the basis of the monoclinic fergusonite structure (SG: *I2/a*) [45] up to the highest pressure reached in our experiments, discarding the possibility of a second phase transition up to 25 GPa. The fact that the high-pressure phase of $SrMoO_4$ has a fergusonite-type structure is in good agreement with previous studies in other alkaline-earth molybdates [28, 29]. It is also in good agreement with the systematics of the high-pressure sequence established for $ABX_4$ compounds using the phase diagram proposed by Bastide [46, 47]. This picture of the structural behavior of $SrMoO_4$ is also consistent with the fact that pressure induces a *sp-d* electron transfer in alkaline-earth metals which convert them into an early transition metal-like character [48, 49]. Therefore, under compression Sr will take an electron configuration similar to that of Y, and it is known that $ABO_4$ ternary oxides of Y and transition metals crystallize in the fergusonite structure [50]. In addition, the existence of group-subgroup relationship between the *I4$_1$/a* and *I2/a* space groups makes the reported transition quite reasonable from the crystallochemical point of view [51].

Fig. 3 shows the Rietveld refinements to the experimental spectra of $SrMoO_4$ at 13.1 GPa obtained assuming the fergusonite structure. In order to perform the Rietveld refinement the starting Sr, W, and O positions have been derived from the atomic coordinates in the scheelite structure using the *I4$_1$/a* → *I2/a* subgroup relationship implemented in the POWDERCELL program [38]. For all the diffraction patterns measured beyond 11.3 GPa, we obtained good agreement between the fergusonite refined model and the experimental diffraction patterns. The full Rietveld refinement of the profile at measured at 13.1 GPa with the fergusonite model converged to small R factors: $R_{WP}$ = 2.25 %, $R_P$ = 1.6 %, and $R(F^2)$ = 1.95 % (198 reflections). Similar refinement quality was obtained for fergusonite $SrMoO_4$ up to 25 GPa. Table I summarizes the lattice parameters and atomic positions obtained for fergusonite $SrMoO_4$. In addition to fergusonite, other structures previously considered as candidates for the high-pressure phases of scheelite-type $ABO_4$ compounds were considered when analyzing the diffraction patterns of the high-pressure phase. These structures were: $BaWO_4$-II (SG: *P2$_1$/n*) [52], $HgWO_4$-type (SG: *C2/c*) [53], raspite (SG: *P2$_1$/a*) [54], wolframite (SG: *P2/c*) [55], α-$MnMoO_4$ (*C2/m*) [17], $LaTaO_4$ (SG: *P2$_1$/c*) [56], $BaMnF_4$ (SG: *A2$_1$/am*) [57], $SrUO_4$ (SG: *Pbcm*), [58], silvanite (SG: *Cmca*) [11], zircon (SG: *I4$_1$/amd*) [59], and pseudo-scheelite (SG: *Pnma*) [60]. However, none of these structures can satisfactorily explain the x-ray diffraction patterns we collected beyond 11.3 GPa. Therefore, it seems quite reasonable to accept that the high-pressure phase of $SrMoO_4$ has a fergusonite structure. Another fact we would like to remark here is that in our experiments we did not find any diffraction peak that could be assigned to the probable decomposition products of $SrMoO_4$ (i.e. SrO and $MoO_3$) [61]. This fact and the reversibility of the high-pressure phase to the scheelite phase on release of pressure imply that $SrMoO_4$ like $SrWO_4$ [11] and $BaMoO_4$ [29] does not decompose significantly into the component oxides at high pressures, unlike what was reported in a previous high-pressure study on $SrWO_4$ [62].

Fig. 4 shows the lattice parameters of the fergusonite phase of $SrMoO_4$ as a function of pressure up to 25 GPa. Experiments were not extended at higher pressure because at 25 GPa the quality of the ADXRD patterns deteriorated. A similar fact was observed previously in SrWO$_4$ [11] and in similar compounds, being independent of the pressure-transmitting medium employed in the experiments. This observation may be related to precursor effects either of a martensitic transition [63 - 65] or of the amorphization observed in alkaline-earth tungstates [16] at higher pressures. In Fig. 4, it can be seen that after the scheelite-fergusonite phase transition the β angle gradually increases with pressure, changing from 90.9° at 13.1 GPa to 92.7° at 25 GPa. On top of that, the difference between the *b/a* and *b/c* axial ratios of the fergusonite phase also increases upon compression; see Fig. 4. These two facts imply an increase of the monoclinic distortion with pressure. In Fig. 5, it can be seen that a volume discontinuity is not apparent at the transition pressure, consistent with the fact that fergusonite is a distorted and compressed version of scheelite which only implies a lowering of the point-group symmetry from *4/m* to *2/m*. A third-order Birch-Murnaghan fit to both the scheelite and the fergusonite pressure-volume data shown in Fig. 5 gives EOS parameters that differ by less than one standard deviation from those obtained for the scheelite data only. Hence, the EOS reported above can be assumed as a valid EOS for SrMoO$_4$ up to 25 GPa, as illustrated in Fig. 5. There it can be seen that the differences between the EOS obtained only the scheelite data (solid line) and that obtained using all the data (dotted line) can be neglected.

Now we would like to comment on the changes we observed in the bond distances at the phase transition. In Fig. 6 it can be seen that at the Mo-O bond distances split, becoming the MoO$_4$ tetrahedra distorted with two short distances (1.739 Å) and two long distances (2.114 Å). In addition to that, the SrO$_8$ dodecahedra distort in such a way that after the transition there are four different Sr-O distances. In spite to these facts, it is interesting to see that in the ferguson ite phase, again, the Sr-O bonds are much more compressible than the Mo-O bonds. As a consequence of this fact, there are virtually no changes between the bulk compressibility of $SrMoO_4$ between its two phases, as can be seen that the volume versus pressure data of both phases can be represented with the same EOS. Basically, in the high-pressure phase the linear compressibility of the different axis changes due to the monoclinic distortion of the crystal, but the bulk compressibility does not change since it is still governed only by the compression of the $SrO_8$ dodecahedra. Another interesting fact we would like to point out here is that at the phase transition two of the oxygen atoms which are second neighbors to Mo approach Mo considerably. In addition, this bond distance rapidly decreases with pressure after the phase transition; see Fig. 6. Therefore, the Mo-O coordination gradually changes from 4 to 4+2 within the ferguson ite phase, as already observed in $PbWO_4$ [10], where it acts as a bridge phase between the scheelite phase and a second high-pressure phase with W-O coordination equal to six.

**C- Structural model for the scheelite-to-fergusonite phase transition**

The pressure-driven transition from the tetragonal scheelite to the monoclinic fergusonite phase has been reported to occur not only in other molybdates [11, 12, 15] and tungstates [28, 29] but also it can be temperature-induced in compounds like $LaNbO_4$ around 780 K [40]. There is also evidence that the transition is of second order [66]. This is consistent with viewing the transition as a slight displacement of the atoms, rather than a more dramatic reconstruction of the lattice. As observed in $BaMoO_4$ [29], in $SrMoO_4$ the scheelite-to-fergusonite transition is caused by small displacements of the Sr and Mo atoms from their high-symmetry positions and large changes in the O positions which consequently lead to a polyhedral distortion (see Fig. 1). In particular, all the Sr and Mo atoms of alternate layers of the scheelite structure shift in opposite directions along the *c*-axis (*b*-axis of the fergusonite structure) accompanied by a shear distortion perpendicular to the *c*-axis of alternate O planes. Because of these atomic displacements, immediately after the transition the volume of $MoO_4$ tetrahedra is enlarged by a factor of 10% and the volume of the $AO_8$ bisdisphenoids is reduced by a similar amount. It is interesting to note that immediately after the transition the fergusonite structure contains isolated $MoO_4$ tetrahedra interlinked by Sr ions which have primarily an eightfold O coordination, like the scheelite structure.

After the phase transition, upon further compression of the fergusonite phase, one pair of parallel unit-cell edges contract, another pair elongates, while the angle between them gradually increases as described above (see Fig. 4). There are two possible ways to achieve this situation and these options can be illustrated by two choices of direction in the tetragonal cell. The two choices are crystallographically identical, and related through the fourfold rotation symmetry of the tetragonal system. In Fig. 8 the transition from the tetragonal system with point group *4/m* to the monoclinic system with point group *2/m* is schematically illustrated and the two orientation states for a transformation from scheelite to fergusonite can be visualized. We will call these two monoclinic orientation states $S_1$ and $S_2$. They are identical in structure, but different in orientation. These orientation states are crystallographically and energetically equivalent. This makes it impossible to distinguish one from the other if they appear separately. All these facts, strongly suggest that the scheelite to fergusonite transition not only is a second-order transformation but it has also a ferroelastic nature. One possibility to probe this hypothesis is to analyze the spontaneous strains of the monoclinic phase, calculated based upon our x-ray diffraction data, using the Landau theory [67]. This analysis is presented in detail in the next section of the paper.

**D- Spontaneous strain and the ferroelastic nature of the phase transition**

In a ferroelastic transformation the $S_1$ and $S_2$ states can be seen as a small distortion caused by slight displacements of the atoms of the parent phase. The spontaneous strain characterizes the distortion of each orientation state relative to the prototype structure (i.e. the scheelite-type structure). Following Schlenker *et al.* [68] the elements of the strain tensor for a crystal can be calculated based upon the lattice parameters. In the case of the tetragonal to monoclinic transition of SrMoO$_4$ the elements of one of the orientation states are:

$$\varepsilon_{11} = \frac{c_M \sin \beta_M}{a_T} - 1 \quad (1),$$

$$\varepsilon_{22} = \frac{a_M}{a_T} - 1 \quad (2),$$

$$\varepsilon_{33} = \frac{b_M}{c_T} - 1 \quad (3),$$

$$\varepsilon_{12} = \varepsilon_{21} = \frac{1}{2} \frac{c_M \cos \beta_M}{a_T} \quad (4),$$

where the subscripts T and M refers to the tetragonal and monoclinic phases. The remaining tensor elements are reduced to zero by the cell parameters. The strain tensor for a single orientation state ($S_1$) is then:

$$e_{ij}(s_1) = \begin{pmatrix} \varepsilon_{11} & \varepsilon_{12} & 0 \\ \varepsilon_{12} & \varepsilon_{22} & 0 \\ 0 & 0 & \varepsilon_{33} \end{pmatrix} \quad (5)$$

and $e_{ij}(S_2)$ is related to $e_{ij}(S_1)$ by $e_{ij}(S_2) = \mathbf{R}\, e_{ij}(S_1)\, \mathbf{R^T}$, where $\mathbf{R}$ and $\mathbf{R^T}$ are the 90° rotation matrix around the *b*-axis of the monoclinic unit cell and its transpose. According to Aizu [69] in the present case the spontaneous strain tensor can be expressed as:

$$e_{ij}^s(s_1) = e_{ij}(s_1) - \frac{1}{2}\left[e_{ij}(s_1) + e_{ij}(s_2)\right] = \begin{pmatrix} -u & v & 0 \\ v & u & 0 \\ 0 & 0 & 0 \end{pmatrix} \quad (6)$$

$$\text{and} \quad e_{ij}^s(s_2) = \begin{pmatrix} u & -v & 0 \\ -v & -u & 0 \\ 0 & 0 & 0 \end{pmatrix} \quad (7),$$

where $u = \frac{1}{2}(\varepsilon_{22} - \varepsilon_{11})$ is the longitudinal spontaneous strain and $v = \varepsilon_{12}$ is the shear spontaneous strain. The scalar spontaneous strain $\varepsilon_s$ is defined as [20, 66, 69]:

$$\varepsilon_s^2 = \sum_{i=1}^{3}\sum_{j=1}^{3}(\varepsilon_{ij}^s)^2 = 2(u^2 + v^2) \quad (8).$$

Following eq. (8) we calculated the spontaneous strains tensor as well as the scalar spontaneous strain for SrMoO$_4$ using the lattice parameter pressure dependences given in Fig. 4. The values of $a_T$ and $c_T$ were extrapolated into the pressure regime of the fergusonite phase (P > 11.3 GPa) from its pressure dependence at pressures lower than the transition pressure. These extrapolations are shown in Fig. 4. We have enough experimental data points within the pressure stability range of the scheelite structure for making a good extrapolation. The obtained results are plotted in Fig. 9 as a function of $\sqrt{(P-P_T)/P_T} = \eta'$, where we choose the transition pressure ($P_T$) as the average of the transition pressure reported here and in Ref. [25], $P_T$ = 12.65 GPa.

The deviation of the fergusonite structure from the $I4_1/a$ symmetry can be expressed by the magnitude of the order parameter η. According to the Landau theory [67] for a second-order transition η is small close to the critical value of the relevant thermodynamic variable (i.e. $P_T$). Under this hypothesis, the Gibbs free energy (G) of the fergusonite phase relative to the scheelite phase can be expressed as a Taylor expansion in terms of η, yielding the following relation if only the first two terms are considered: $G = k_1(P-P_T)\eta^2 + k_2\eta^4$, where $k_1$ and $k_2$ are two constants [20]. As pressure drives the studied transition, from this equation, the relation between pressure and the order parameter can be found by minimizing G; i.e. when the condition $\partial G / \partial \eta = 0$ is satisfied. This condition is only fulfilled if the order parameter has the form: $\eta \propto \sqrt{(P-P_T)/P_T} = \eta'$, which can be defined as the phenomenological order parameter in Landau's theory [67]. In a ferroelastic transition $\varepsilon_s$ can be considered as being proportional to the primary order parameter η [70] and consequently also to η'. Therefore, if the pressure-induced scheelite to fergusonite phase transition is a second-order ferroelastic transition $\varepsilon_s$ should be a linear function in Fig. 9. In this figure, it can be seen that, within the uncertainty of the results, this condition is satisfied in SrMoO$_4$. Indeed, we found that $\varepsilon_s$ = 0.0585(8) η', with a correlation coefficient $r^2$ = 0.994. This fact strongly suggests that the scheelite to fergusonite transition studied here is a second-order phase transition. An analogous phase transition was obtained after analyzing the spontaneous strain in alkaline-earth orthotungstates [20] and in that case the proportionality constant was 20% smaller than in the present investigation. Further, a ferroelastic transition has been also found at low temperature in the isostructural scheelite $CaMoO_4$ [71]. In addition, the observation of a soft acoustic mode in Brillouin scattering measurements in scheelite-structured $BiVO_4$ [72] is another conclusive evidence of the correction of the ferroelastic interpretation of the scheelite to fergusonite transition. All these facts give additional support to our conclusion.

## IV. Conclusions

Our x-ray diffraction studies on strontium molybdate show that it transforms to the fergusonite phase around 12.2(9) GPa, as also observed in isostructural barium molybdate [29] and strontium tungstates [17]. No additional phase transitions were observed up to 25 GPa and on release of pressure $SrMoO_4$ reverts back to the initial scheelite phase. The high pressure phase is a distorted and compressed version of scheelite obtained by a small distortion of the cation matrix and significant displacements of the anions. A mechanism for the transition is proposed and its character is discussed using the calculated spontaneous strain and the Landau theory. In particular, we found that the spontaneous strain is a linear function of the phenomenological order parameter defined in the Landau theory. This evidence strongly suggests that the studied pressure-induced phase transition is a second-order ferroelastic phase transformation. Finally from the EOS of $SrMoO_4$ we have obtained $B_0$ = 71(3) GPa, $B_0'$ = 4.2(4) with $V_0$ = 349.7(4) $Å^3$. The compressibility of both the low-pressure and high-pressure phase is highly anisotropic. This fact and the determined bulk compressibility are discussed in terms of the different compressibility of the $SrO_8$ and $MoO_4$ polyhedra.

**Acknowledgments**

This study was made possible through financial support from the Spanish government MCYT under Grant No. MAT2004-05867-C03-01 and of the Nature Science Foundation of the Fujian Province of China under Grant No. 2005HZ1026. The U.S. Department of Energy, Office of Science, and Office of Basic Energy Sciences supported the use of the APS under Contract No. W-31-109-Eng-38. DOE-BES, DOE-NNSA, NSF, DOD-TACOM, and the Keck Foundation supported the use of the HPCAT. D.E. acknowledges the financial support from the MCYT of Spain and the University of València through the "Ramón y Cajal" program. Work at UNLV is supported by DOE award No. DEFG36-05GO08502. The UNLV High Pressure Science and Engineering Center is supported by the DOE-NNSA under cooperative agreement No. DE-FC52-06NA26274.

**Table I:** Structural parameters of the scheelite and fergusonite structure of $SrMoO_4$. These parameters were obtained from the present Rietveld refinements (see text).

(a) Structural parameters of scheelite $SrMoO_4$ at 0.5 GPa: $I4_1/a$, Z = 4, $a$ = 5.380(5) Å, $c$ = 12.019(9) Å.

|    | Site | x        | y        | z        |
|----|------|----------|----------|----------|
| Sr | 4b   | 0        | 0.25     | 0.625    |
| Mo | 4a   | 0        | 0.25     | 0.125    |
| O  | 16f  | 0.248(9) | 0.110(6) | 0.051(6) |

(b) Structural parameters of fergusonite $SrMoO_4$ at 13.1 GPa: $I2/a$, Z = 4, $a$ = 5.256(5) Å, $b$ = 11.191(9) Å, $c$ = 5.195(5) Å, $\beta$ = 90.9(1)°.

|    | Site | x        | y        | z        |
|----|------|----------|----------|----------|
| Sr | 4e   | 0.25     | 0.615(2) | 0        |
| Mo | 4e   | 0.25     | 0.123(1) | 0        |
| O  | 8f   | 0.910(6) | 0.951(6) | 0.248(5) |
| O  | 8f   | 0.492(6) | 0.239(5) | 0.797(6) |

**Figure Captions**

**Figure 1: (color online)** (a) The scheelite structure of SrMoO$_4$ at 0.5 GPa and (b) the ferguosnite structure of SrMoO$_4$ at 24 GPa. Large blue circles represent the Sr atoms, middle-size blue circles the Mo atoms and small black circles the O atoms. The unit cell, Sr-O bonds, and Mo-O bonds are also shown. In (b) we also illustrate that at high pressure the Mo coordination changes from 4 to 6 with two short bonds (gray), two middle-size bonds (red), and two large bonds (black); see text and Fig. 6.

**Figure 2:** Room-temperature ADXRD data of SrMoO$_4$ at different pressures up to 24 GPa. In all diagrams the background was subtracted. Pressures are indicated in the plot.

**Figure 3:** ADXRD pattern of SrMoO$_4$ at 0.5 GPa and 13.1 GPa. The background was subtracted. The dots are the experimental data and the solid lines the refined profiles. The dotted lines represent the difference between the measured data and the refined profiles. The bars indicate the calculated positions of the reflections.

**Figure 4:** Evolution of the lattice parameters and axial ratios SrMoO$_4$ under pressure. Circles: scheelite phase. Squares: ferguosnite phase. An extrapolation of the lattice parameters of the low-pressure phase up to 25 GPa is also shown.

**Figure 5:** Volume versus pressure data for scheelite (circles) and ferguosnite (squares) SrMoO$_4$. The solid line represents the EOS fitted using only the scheelite data and the dotted line the EOS fitted using all the data.

**Figure 6:** Pressure dependence of the interatomic bond distances in the scheelite and fergusonite phases of SrMoO$_4$. Squares represent the Mo-O distances and circles the Sr-O distances.

**Figure 7:** *c/a* axial ratio of SrMoO$_4$ as a function of the atomic volume. Circles: present high-pressure data. Squares: data corresponding to the volume expansion observed at high temperature [42]. The solid line is just a guide to the eye. The dashed line delimitates the compression and expansion regions.

**Figure 8:** A schematic view of two orientation states of the plane perpendicular to the *c*-axis for a scheelite to fergusonite transformation. Because of the fourfold-symmetry of the prototypic phase there are two possible ways for the unit cell to deform. $\varepsilon$ and $\delta$ represent the contraction or expansion of the tetragonal *a*- and *b*-axis.

**Figure 9:** Correlation between the spontaneous strain $\varepsilon_s$ and the Landau order parameter $\eta'$. Squares: experimental data. Line: least-squares linear fit.

**Figure 1**

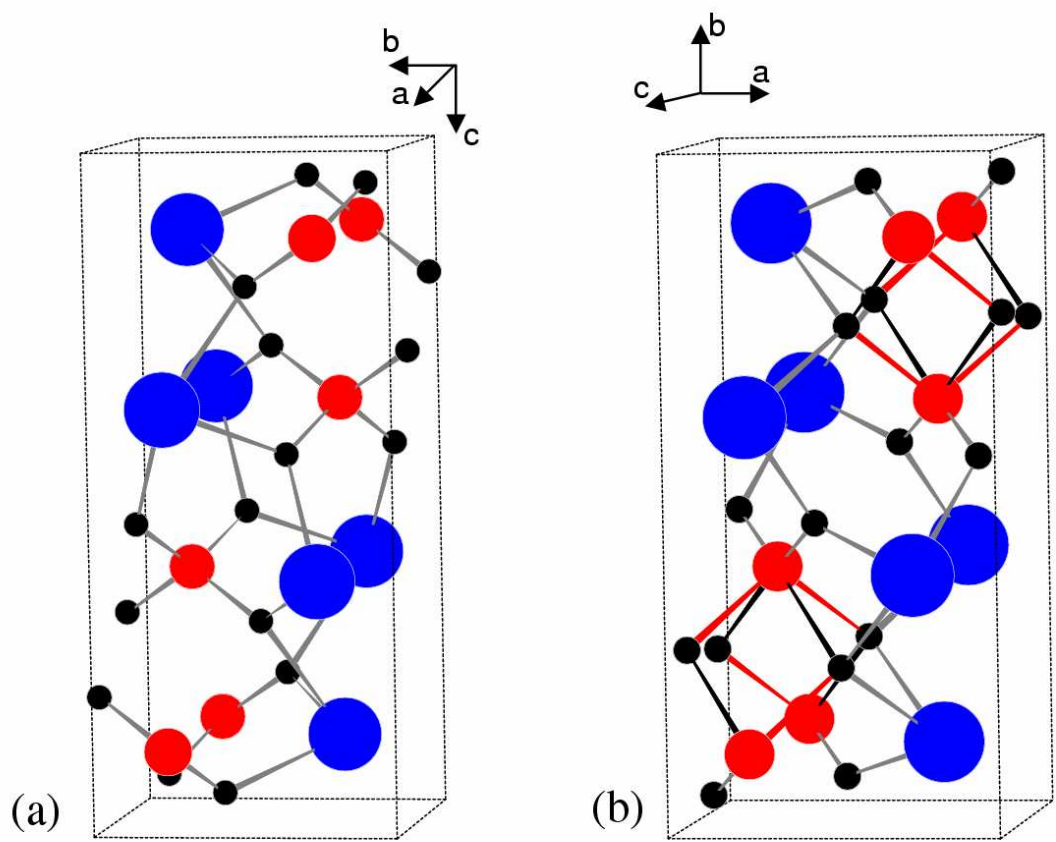

**Figure 2:**

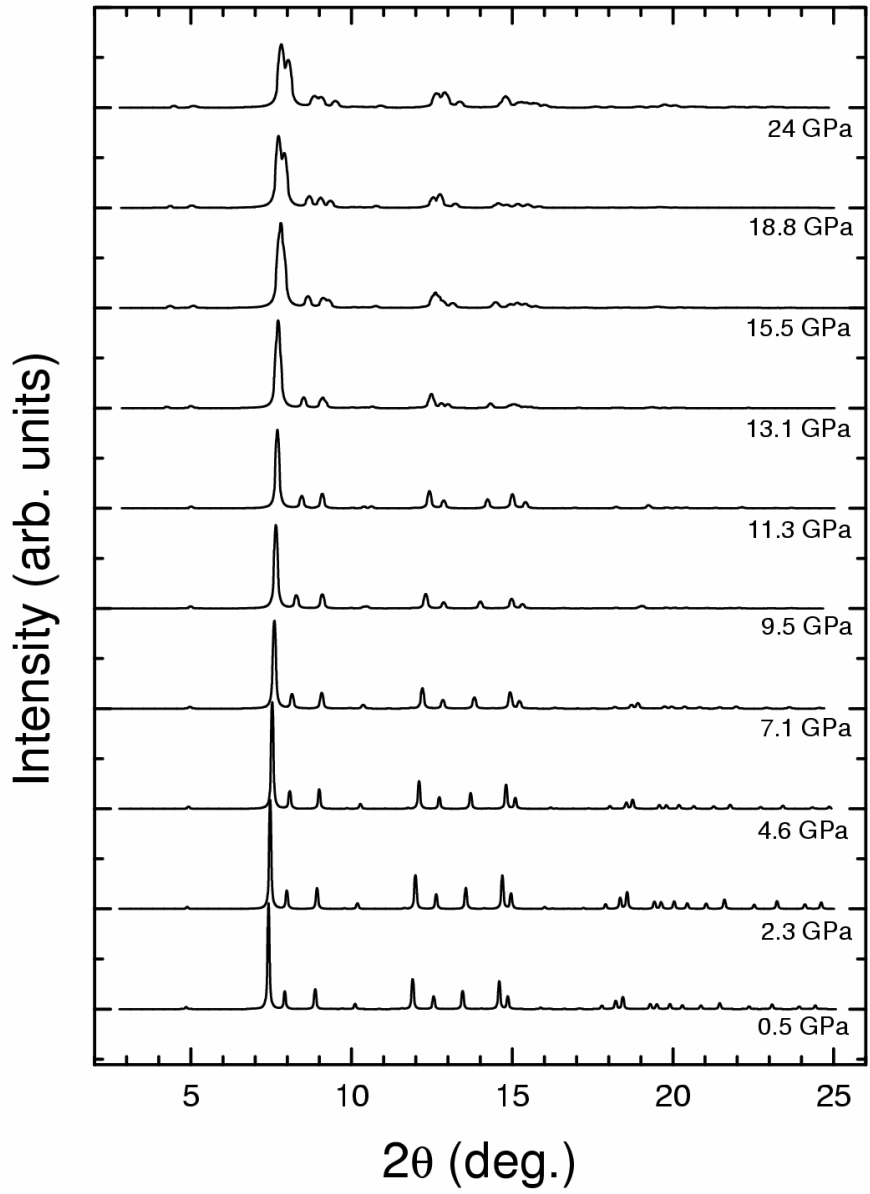

**Figure 3:**

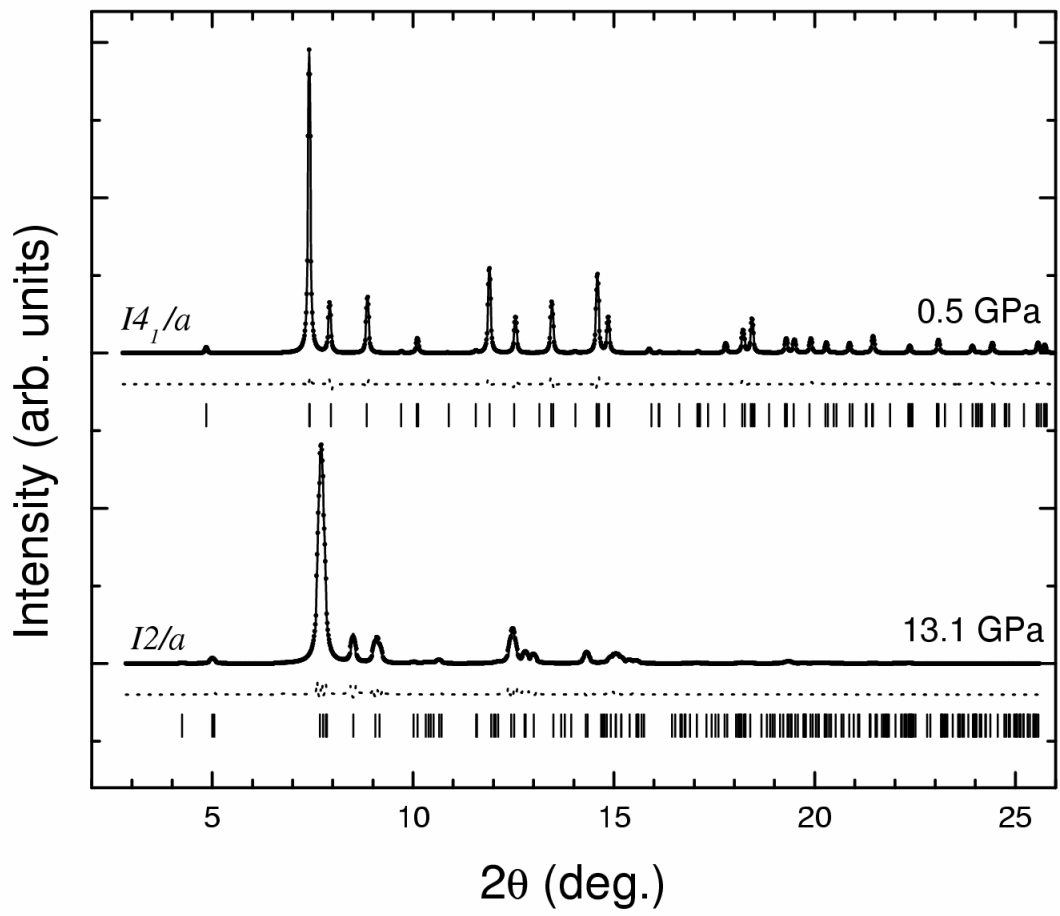

**Figure 4:**

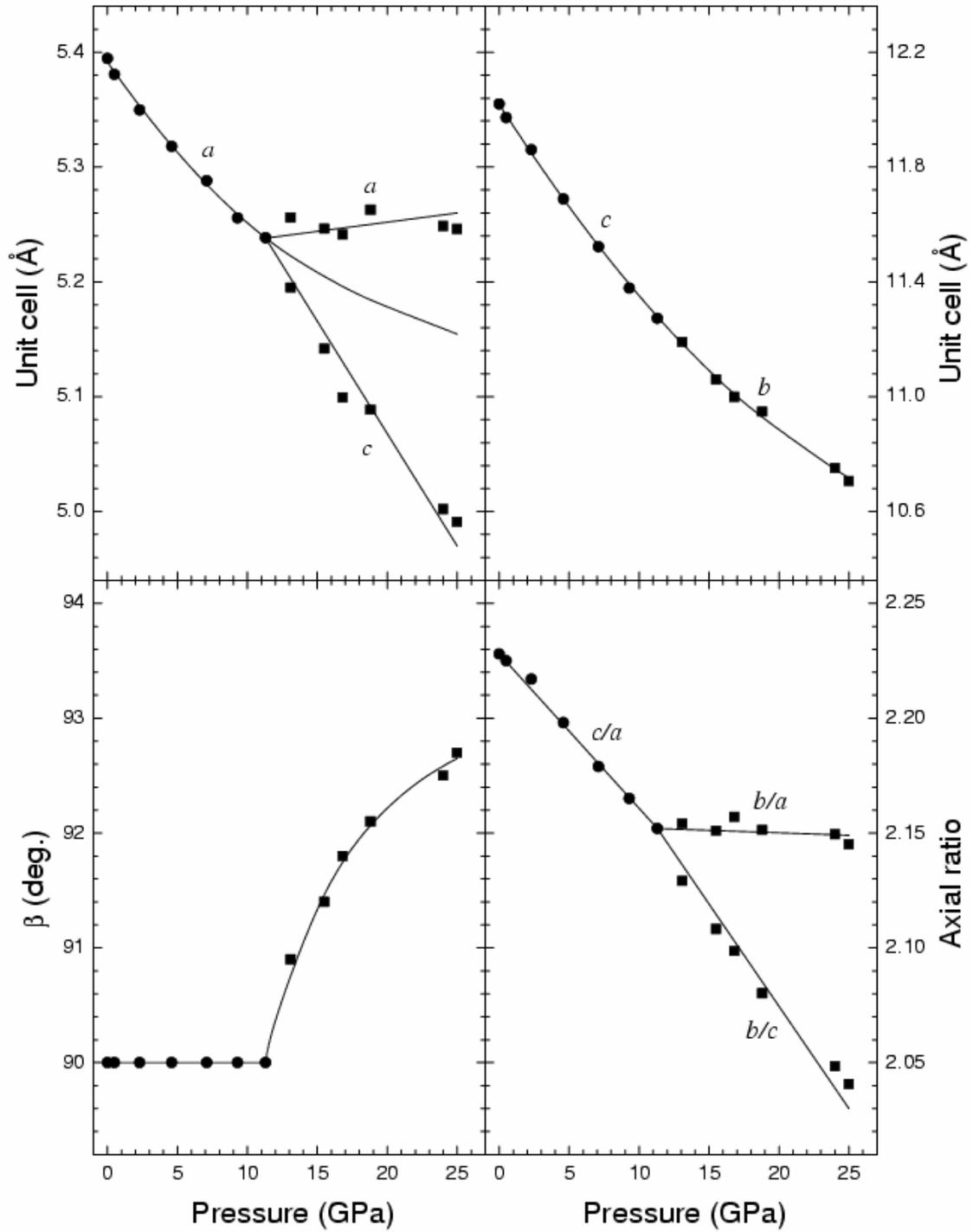

**Figure 5:**

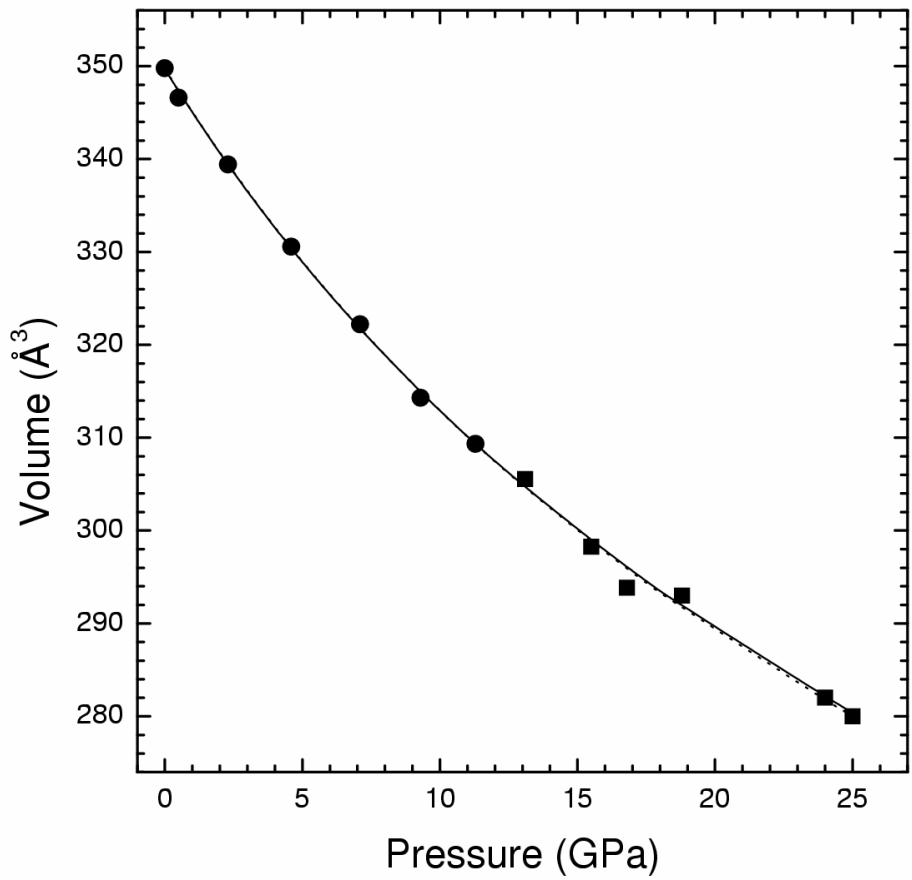

**Figure 6:**

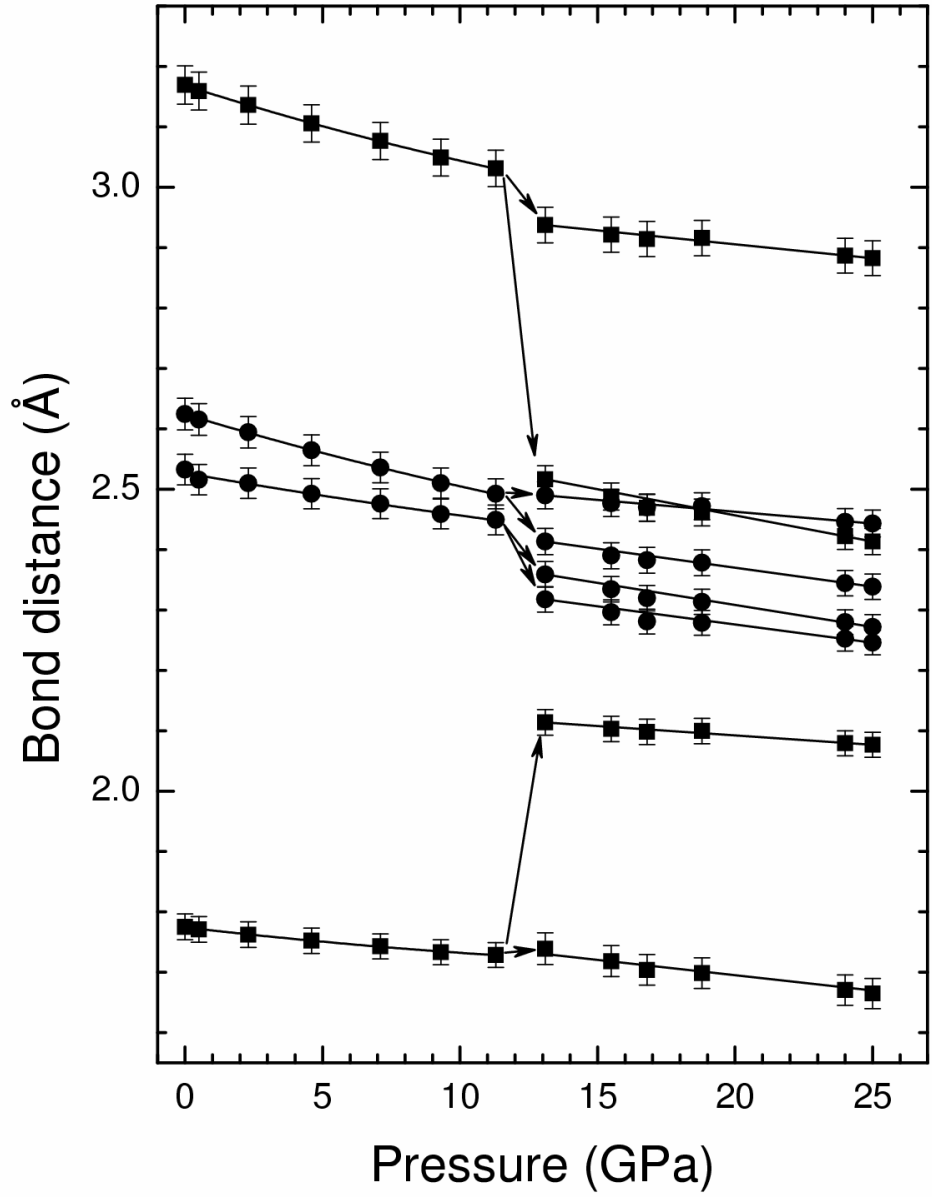

**Figure 7:**

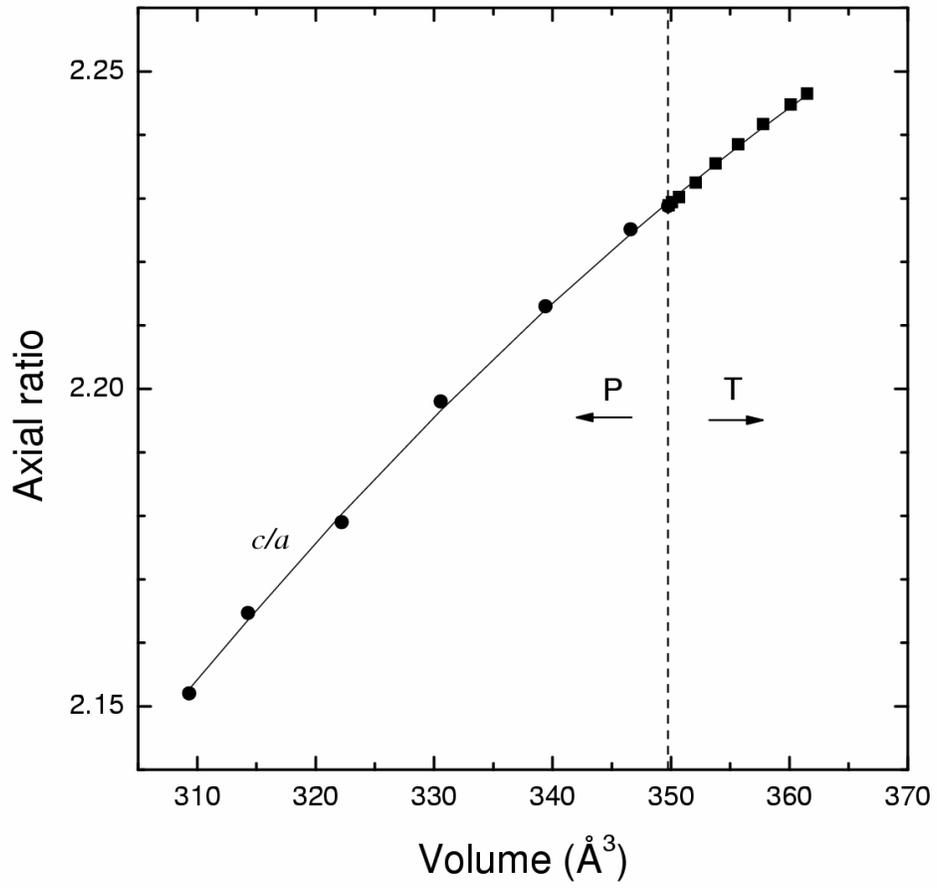

**Figure 8:**

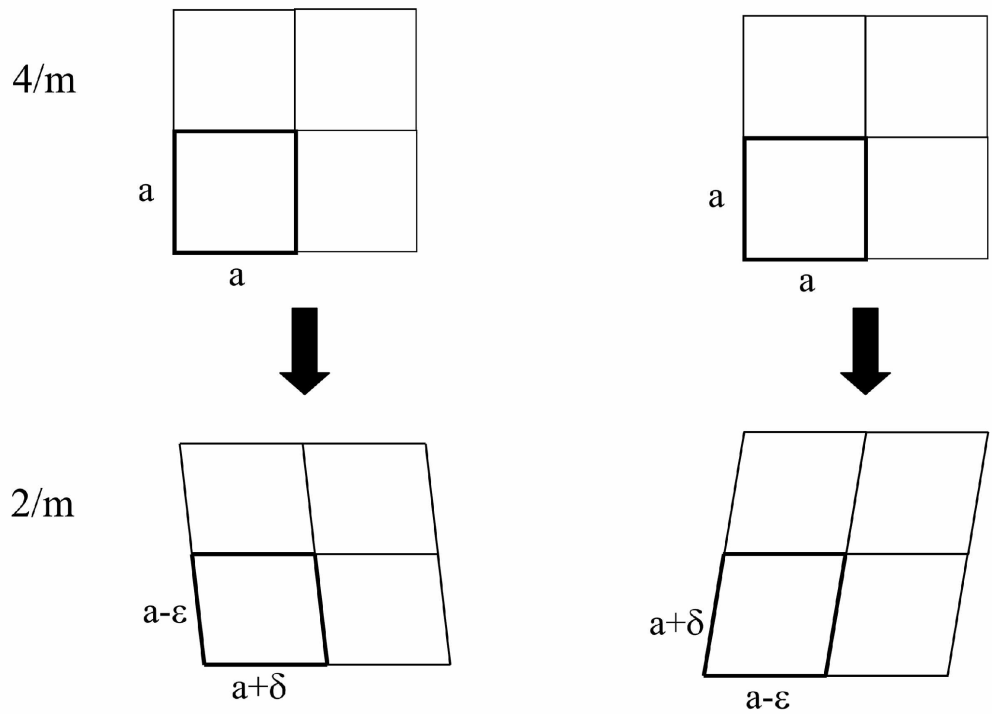

**Figure 9:**

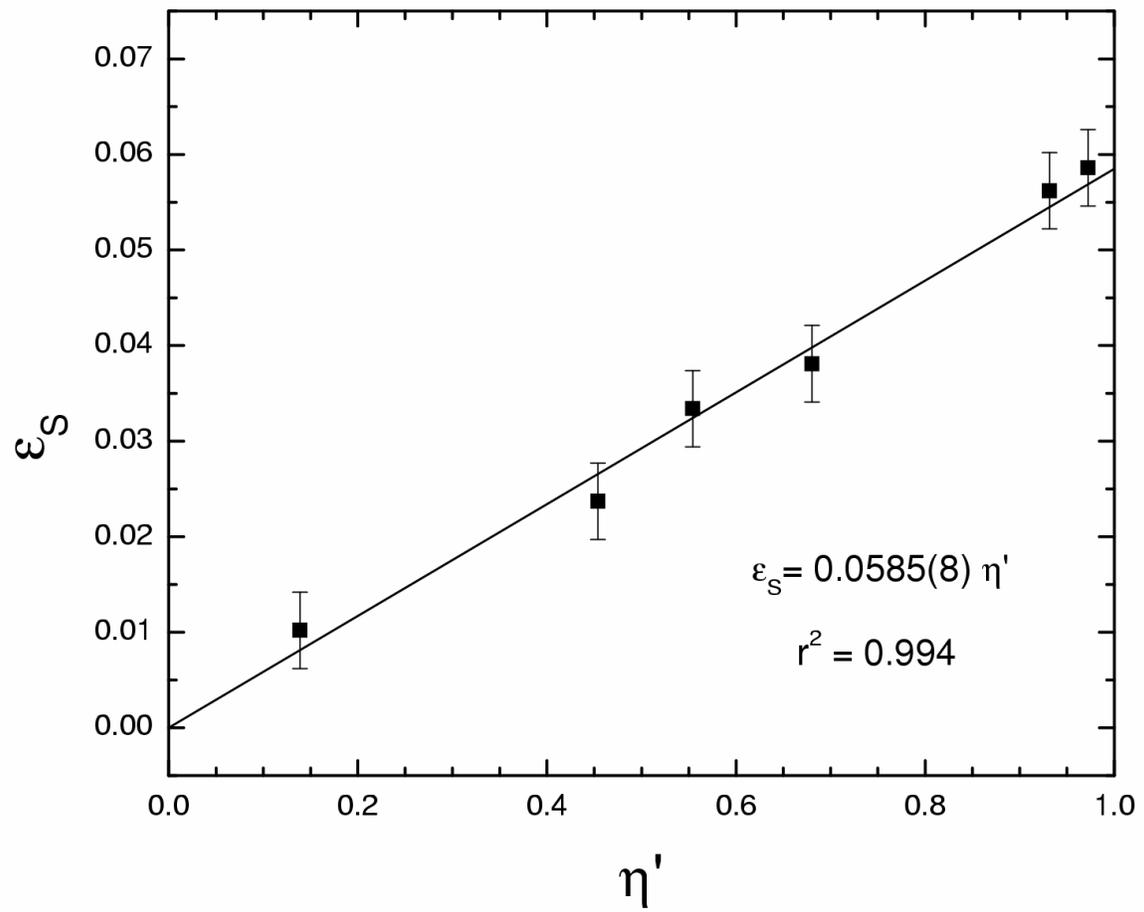